\documentclass[a4paper,10pt]{article}
\usepackage{amssymb}
\usepackage{amsmath}
\usepackage{multirow}
\usepackage{graphicx}
\usepackage{geometry} 

\newgeometry{tmargin=2.5cm, bmargin=1.5cm} 
\def\d#1{{\rm d}#1}
\def\calD{{\cal D}}
\def\calI{{\cal I}}
\def\calK{{\cal K}}

\def\calM{{\cal M}}
\def\calP{{\cal P}}
\def\mitPhi{ \mathit{\Phi}}

\begin{document}

\title{\bf First integrals in the Brans--Dicke cosmology}
\author{ Zdzis{\l}aw A. Golda$^{1,3}$, Andrzej Woszczyna\thanks{uowoszcz@cyf-kr.edu.pl}\,\,\,$^{2,3}$, {\L}ukasz Bratek$^{2}$}
\date{}
\maketitle

\vspace{-10mm}

\begin{center}
{{$^1$}Astronomical Observatory, Jagiellonian University,\\  ul. Orla 171, 30--244 Krak\'ow, Poland}\\
\smallskip
{{$^2$}Institute of Physics, Cracow University of Technology,\\ ul. Podchor\c{a}\.zych 1, 30--084 Krak\'ow, Poland}\\
\smallskip
{{$^3$}Copernicus Center for Interdisciplinary Studies,\\ ul. S{\l}awkowska 17, 31--016 Krak\'ow, Poland}
\end{center}
\bigskip
\bigskip 

\begin{abstract}
\noindent 
The work presents the first and the second degree Darboux polynomials, Jacobi's last multipliers as well as the set of first integrals for Brans--Dicke cosmology. Algebraic invariant sets are constructed. First integrals are visualized for some particular values of the $\omega$ parameter. 
\end{abstract}


\bigskip

\def\hang{\hangindent\parindent}
\def\textindent#1{\indent\llap{#1\enspace}\ignorespaces}
\def\litem{\par\hang\textindent}
\def\subitem{\par\indent \hangindent2\parindent\textindent}
%
%
%
%

\section{Introduction}
\label{sec:01}
Brans--Dicke gravity theory assumes that in addition to the metric tensor  $g_{\mu\nu}$  there is a long-range scalar field  $\varphi$ that determines the gravitational ``coupling constant''. Alike the metrics  $g_{\mu\nu}(t,x^a)$,  the additional  field  $\varphi(t,x^a)$ is a space and time dependent quantity. It is shaped by the distribution of  matter in the universe. The complete field equations for this theory are  given in~\cite{01}, details and physical consequences have been discussed in~\cite{02}. Brans--Dicke gravity theory belongs to the canon of cosmology~\cite{03}, and the theory is presently considered as a possible explanation of  the dark energy phenomenon.

In the present paper we investigate constants of motion for the expanding, homogeneous and isotropic space-time with varying gravitational coupling factor. We employ Darboux  techniques to find first integrals. We limit ourselves to the early stage cosmology i.e. to barotropic matter with the equation of state $p=\rho/3$. In the section  \ref{sec:02} and \ref{sec:03}   we introduce dynamical equations for Bans--Dicke cosmology. In the section  \ref{sec:04} we provide Darboux polynomials and algebraically invariant sets. The Appendix tables provide the classification of  algebraically invariant sets for special solutions to Brans--Dicke dynamical system.

\section{ Brans--Dicke scalar-tensor theory of gravity}
\label{sec:02}

Alike in General Relativity, the energy-momentum tensor $ T_{\mu \nu}$ with vanishing divergence $\nabla_\nu T^{\mu\nu}=0$ is the source of  Brans--Dicke gravitational field.
However, in Brans--Dicke theory the trace of the energy-momentum tensor  $T=T_\rho^\rho$ excites an additional scalar field $\varphi$
		\begin{equation}
\Box\,\varphi=\frac{8\pi}{c^4(3+2\omega)}T_\rho^\rho.
		\label{eq:pole_phi}
		\end{equation}
witch, in turn, enters the gravity field equations twice
\arraycolsep=0.15em
		\begin{eqnarray}
R_{\mu\nu}-\frac{1}{2}R\,g_{\mu\nu}&=&\frac{8\pi\varphi^{-1}}{c^4}T_{\mu\nu}+\frac{\omega}{\varphi^2}
\left[
(\nabla_\mu\varphi)(\nabla_\nu\varphi)-\frac{1}{2}g_{\mu\nu}(\nabla_\rho\varphi)(\nabla^\rho\varphi)
\right]\nonumber\\
&&{}+\varphi^{-1}(\nabla_{\mu\nu}\varphi-g_{\mu\nu}\Box\,\varphi).
		\label{eq:tensorR}
		\end{eqnarray}
first, as $\varphi$--dependent coupling factor, and second, as a correction term on the right hand side.

$\omega$ is a free parameter of the theory to be empirically  determined. For a gravitational constant to be positive $0 <G <\infty$ we put $2\omega+3> 0$.
For  $\varphi=\mbox{const}=G^{-1}$ and $\omega\to\infty$  Brans--Dicke theory reduces to  Einstein's general relativity.  Brans--Dicke theory has also generalisations~\cite{04}.

Brans--Dicke theory does not affect the isometry group properties. The maximal isometry group acting on space-like  hypersurfaces leads to Robertson--Walker metric form, 
		\begin{equation}
\d s^2=\d t^2- R^2\left[
\frac{\d r^2}{1-k r^2}+r^2\left(\d \theta^2+\sin^2 \theta \d \varphi^2\right)
\right], 
		\label{eq:metricRW}
		\end{equation}
and to the strictly time dependent scalar field $\varphi=\varphi(t)$.
Thus, for homogeneous and isotropic universe the equations (\ref{eq:pole_phi}) and (\ref{eq:tensorR}) take form
\begin{eqnarray}
\arraycolsep=0.1em
&\varphi''+3\frac{R'}{R}\varphi'=\frac{\rho-3p}{3+2\omega},
\label{eq:kos2}\\
&\frac{R'^2}{R^2}+\frac{k}{R^2}=\frac{\rho}{3\varphi}+\frac{\omega}{6}\frac{\varphi'^2}{\varphi^2}-\frac{\varphi'}{\varphi}\frac{R'}{R},
\label{eq:kos3}\\
&\frac{R''}{R}=-\frac{1}{3(3+2\omega)}
\frac{(3+\omega)\rho+3\omega p}{\varphi}
-\frac{\omega}{\varphi}\frac{\varphi'^2}{\varphi^2}+\frac{\varphi'}{\varphi}\frac{R'}{R}
\label{eq:kos4},\\
&\rho'+3\frac{R'}{R}(\rho+p)=0,
\label{eq:kos1}
		\end{eqnarray}
where $\rho=\rho(t)$ stands for energy density, $p=p(t)$ stands for pressure, $R=R(t)$ is the scale factor,  $k=0,\pm 1$ is the signature of the space curvature. We follow the convention $\frac{8\pi}{c^4}=1$.

\section{The Brans--Dicke dynamical system}
\label{sec:03}

\subsection{System definition}

The system (\ref{eq:kos2}--\ref{eq:kos1}) is differentially dependent, so we choose equations  (\ref{eq:kos2},\ref{eq:kos4},\ref{eq:kos1}) for further analysis. To close the system one has to supplement it by the equation of state of the matter content. In what follows we assume  $p=\alpha\rho$, with  $\alpha=1/3$. This  approximates the matter properties in the epochs  prior to the last scattering.

Under redefinition  of  variables
		\begin{equation}
P=\frac{\rho}{\varphi},~~~~H=\frac{R'}{R},~~~~\mitPhi=\frac{\varphi '}{\varphi }.
		\label{eq:DynamicVariables}
		\end{equation}
one obtains the third order dynamical system
		\begin{eqnarray}
\arraycolsep=0.1em
\left\{ \!\!\!
		\begin{array}{rcl}
		P'&=&f_1:=-3 (1+\alpha)P H-P \mitPhi ,
\label{komutator1}\\[1ex]
		H'&=& f_2:=-\frac{3+(1+3\alpha) \omega}{3 (3+2 \omega
   )}P-H^2+H \mitPhi-\frac{\omega }{3}\mitPhi ^2,
\label{komutator2}\\[1ex] 
		\mitPhi'&=&f_3:=\frac{1-3 \alpha }{3+2 \omega}P-3 H \mitPhi -\mitPhi^2.
\label{komutator3}
		\end{array}
		\right.
		\label{eq:DynamicalSystem}
		\end{eqnarray}
Dynamical variables (\ref{eq:DynamicVariables}) have the following interpretation: 
\litem{$\circ$}$P$ --- density of gravitational mass, 
\litem{$\circ$}$H$ --- expansion rate (Hubble parameter),
\litem{$\circ$}$\mitPhi$ --- field evolution rate.

\noindent Components of the  vector field $f: = (f_1, f_2, f_3)$ of the system (\ref{eq:DynamicalSystem}) are the second degree polynomials in ${P,H,\mitPhi}$.

\subsection{General properties and symmetries}
\label{subsec:symetrie}

For $\alpha=\frac{1}{3}$ the system has the following properties:
\litem{$\circ$}$(0,0,0)$ --- is the only complex critical point of the $(P,H,\mitPhi)$ phase space such that all the eigenvalues of Jacobi matrix vanish,
\litem{$\circ$}nonvanishing divergence of the vector field $\nabla{\cdot} f=-9H - 2 \mitPhi$ means that the system is dissipative,
\litem{$\circ$}system admits the time  reflection $t\longrightarrow -t \Longrightarrow (P,H,\mitPhi)\longrightarrow(P,-H,-\mitPhi) $,
\litem{$\circ$}system has the scaling symmetry $t\longrightarrow \frac{t}{\lambda} \Longrightarrow (P,H,\mitPhi)\longrightarrow(\lambda^2 P,\lambda H,\lambda \mitPhi) $.

\section{Darboux polynomials, first integrals, Jacobi's last multipliers and algebraically invariant sets}
\label{sec:04}

\subsection{Darboux polynomials}

There are no standard  techniques to find first integrals of ordinary differential equations, yet there is a number of special methods widely discussed in the literature. Among them there are methods appealing to  Lie symmetries~\cite{07,08,09}, Noether symmetries~\cite{10}, Lax pairs~\cite{11}, Painlev\'e analysis~\cite{12}, Carleman embedding~\cite{13,14,15,16}, differential Galois theory~\cite{17,18} and Darboux's method~\cite{19,20,21,22}. The Darboux polynomial method is also discussed in a few handbooks~\cite{05,23,06}.

The recently growing interest in the Darboux method is stimulated by fast development of computer algebra languages. Although  strongly limited to the  equations of  polynomial coefficients, this method became attractive due to its algorithmic nature.
Using this approach for the Brans--Dicke system (\ref{eq:DynamicalSystem}) we get five irreducible Darboux polynomials $\calP$ with corresponding cofactors  $\calK$:\\ \\
The first degree Darboux polynomials for the system (\ref{eq:DynamicalSystem}) are:
	\begin{align}
	&\calP_{11}=P,\label{eq:WD01} \\
	&\calK_{11}=-4 H - \mitPhi,
     	\end{align}
	\begin{align}
\label{eq:WD02}
     	&\calP_{12}=\mitPhi,\\
     	&\calK_{12}=-3H-\mitPhi.
     	\end{align}
The second degree Darboux polynomials for the system (\ref{eq:DynamicalSystem}) are:
	\begin{align}
\label{eq:WD03}
	&\calP_{21}=2 P - 6H(H+\mitPhi) + \omega\mitPhi^2,\\
	&\calK_{21}=-2 H,
	\end{align}
	\begin{align}
	&\calP_{22}=2 P+2\sqrt{9+6\omega}H\mitPhi +(3+2\omega+\sqrt{9+6\omega})\mitPhi^2,\\				&\calK_{22}=-4 H-\frac{1}{3}(3+\sqrt{9+6\omega})\mitPhi,
     	\end{align}
	\begin{align}
	&\calP_{23}=2 P-2\sqrt{9+6\omega}H\mitPhi +(3+2\omega-\sqrt{9+6\omega})\mitPhi^2,\\				&\calK_{23}=-4 H-\frac{1}{3}(3-\sqrt{9+6\omega})\mitPhi.
     	\end{align}
By direct calculation we have excluded irreducible third degree Darboux polynomials. The first degree Darboux polynomials  (\ref{eq:WD01}) and  (\ref{eq:WD02}) coincide with the dynamical variables. The second degree polynomial  (\ref{eq:WD03}) is equivalent to the equation of constraint  (\ref{eq:kos3}) for $k=0$.

\subsection{Darboux integrals}
\label{subsec:Darbouxintegrals}

Following the standard procedure we employ  Darboux polynomials $\{\calP_{11}, \calP_{12}, \calP_{21}, \calP_{22}, \calP_{23}\}$ to construct two families of first integrals $ {\cal I}_{1,2}$, i.e functions satisfying $f{\cdot}\nabla {\cal I}_{1,2}=0$.

\begin{figure}[h]
\vspace*{-2.5ex}
		\begin{center}
			\begin{tabular}{cc}
\hspace*{-0.5em}\includegraphics[width=0.48\textwidth]{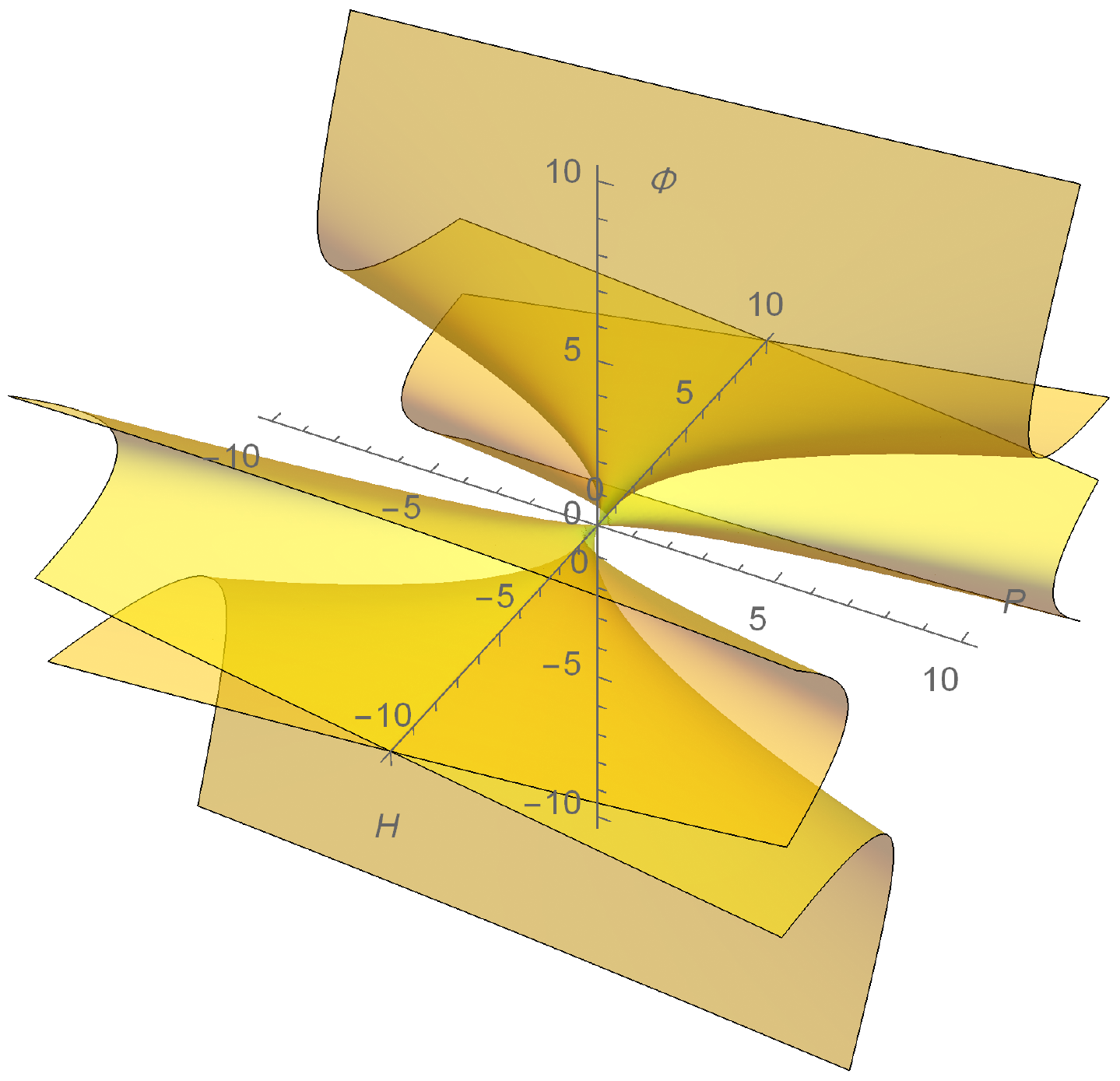}&
\includegraphics[width=0.48\textwidth]{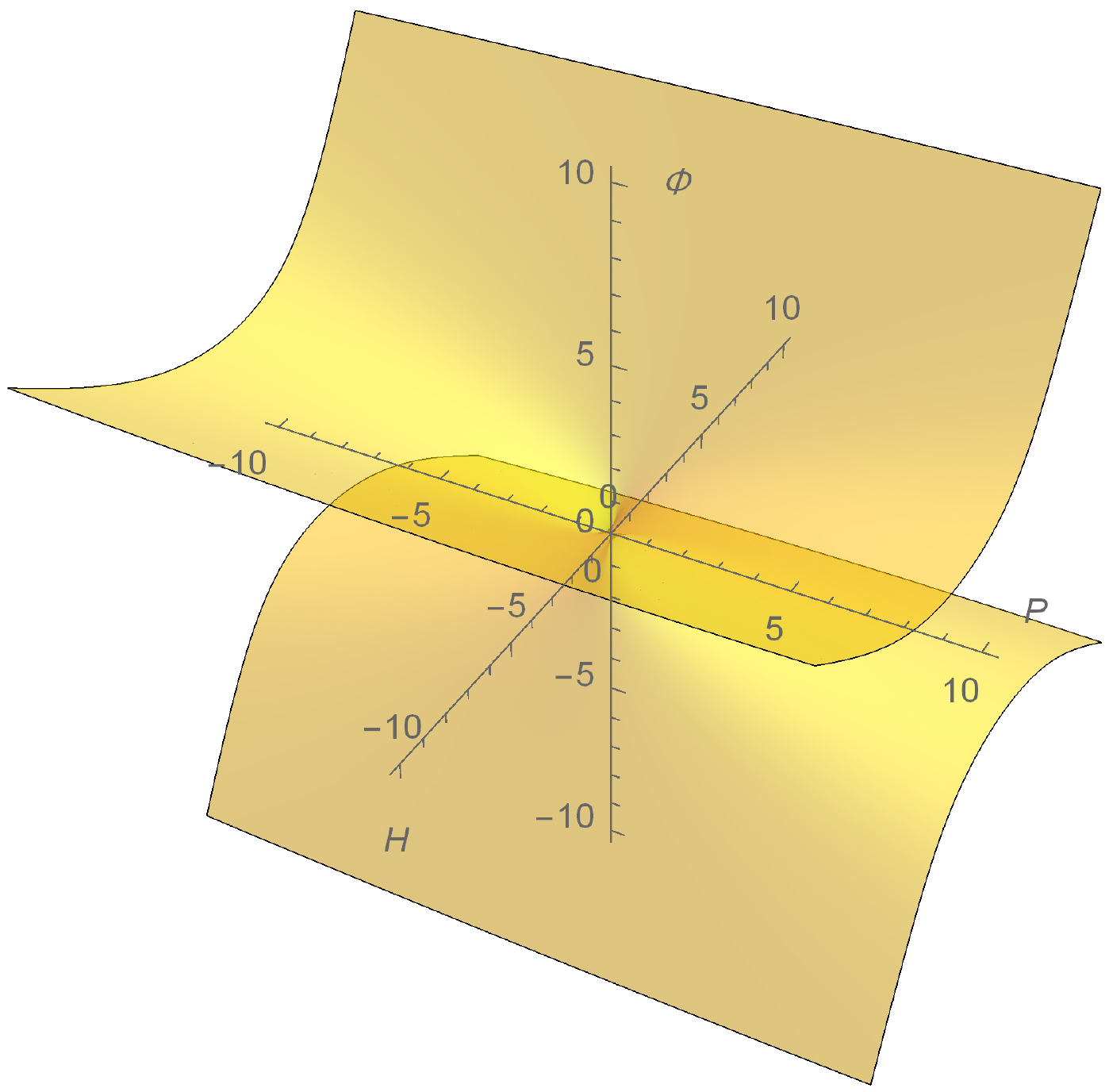}
			\end{tabular}
		\end{center}
\vspace{-4.5ex}
\caption{Illustration of the first integral  ${\cal I}_1$ for $\omega=9/2$ and ${\cal I}_1=-10$ [left panel] and ${\cal I}_1=10$ [right panel].}
\label{Figure01}
\vspace{-1mm}
	\end{figure}

First integral ${\cal I}_1$ is the rational function of variables $\{P,H,\mitPhi\}$ --- homogeneous  of degree zero
		\begin{equation}
{\cal I}_1=\calP_{11}^{-2}\calP_{12}^2\calP_{21}=P^{-2}\mitPhi^{2} (2 P-6 H^2 -  6 H \mitPhi + \omega \mitPhi^2).
		\label{eq:I1}
		\end{equation}
For illustration see Figure~\ref{Figure01}. On the strength of equations  (\ref{eq:kos3}) and  (\ref{eq:DynamicalSystem}), first integral  $\calI_1$  can be expressed as $\calI_1=\gamma k$, where $\gamma\in \mathbb{R}$. This means that  $\calI_1$ is proportional to the space curvature.

\begin{figure}[h]
\vspace*{-2.0ex}
		\begin{center}
			\hspace*{-1em}\includegraphics[width=0.65\textwidth]{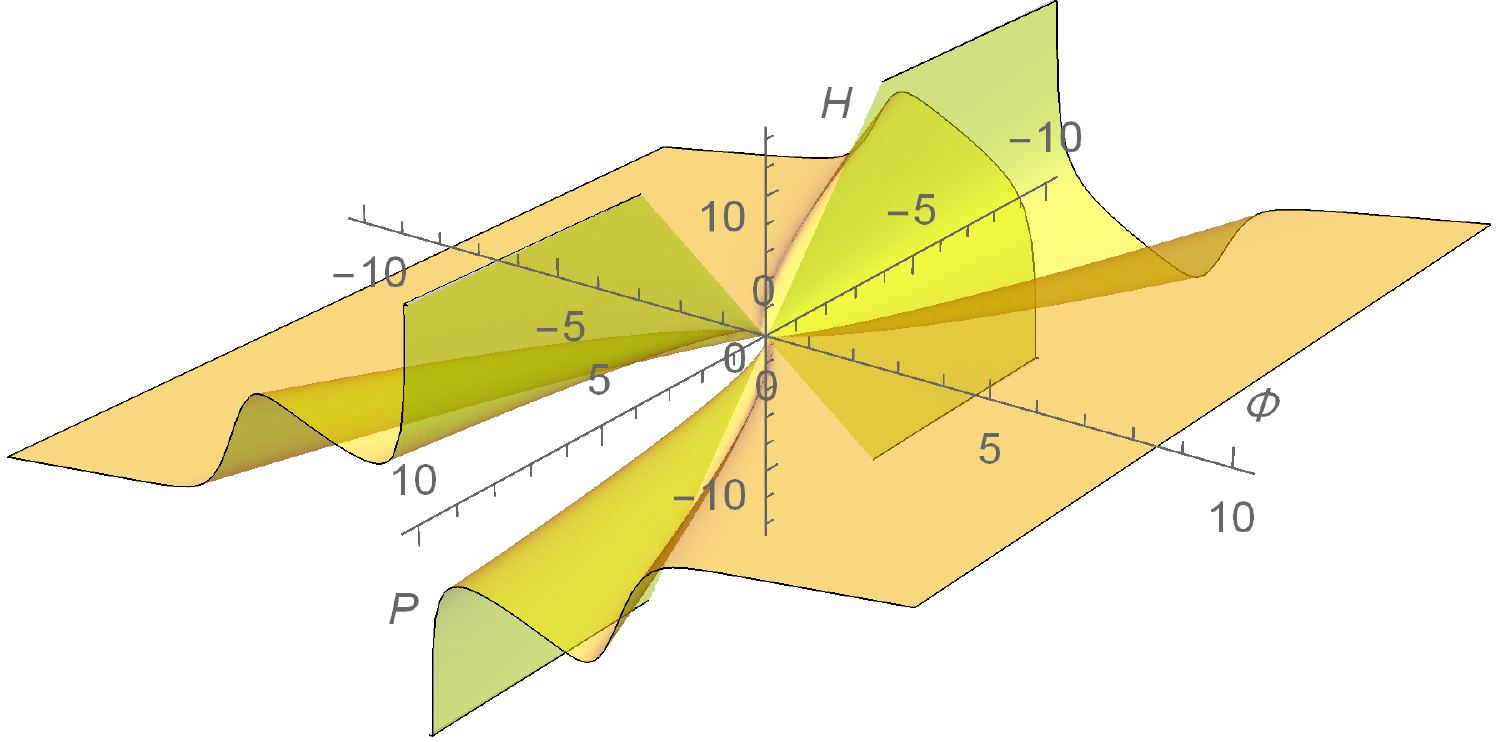}
		\end{center}
\vspace{-3.5ex}
\caption{Illustration of the first  integral ${\cal I}_2$ for $\omega=9/2$ and ${\cal I}_2=10$.}
\label{Figure02}
\vspace{-1mm}
	\end{figure}

First integral  ${\cal I}_2$  is the function  homogeneous of degree four, and can be expressed as
		\begin{equation}
{\cal I}_2=\calP_{11}^6 P_{12}^{-8}
\left[\frac{\calP_{22}}{\calP_{23}}
\right]^\beta=P^6\mitPhi^{-8} 
\left[
\frac{2P+2\sqrt{9+6\omega}H\mitPhi +(3+2\omega+\sqrt{9+6\omega})\mitPhi^2}{2 P-2\sqrt{9+6\omega}H\mitPhi +(3+2\omega-\sqrt{9+6\omega})\mitPhi^2}
\right]^\beta,
		\label{eq:I2}
		\end{equation}
where the exponent $\beta=\frac{\sqrt3}{\sqrt{3+2\omega}}$ (here and later in the article).  ${\cal I}_2$ is determined for $\calP_{22}{}\calP_{23}>0$  and, in general, is not a rational function. See  Figure~\ref{Figure02}.

First integrals ${\cal I}_1$ and ${\cal I}_2$ are functionally independent in the common domain\break  
$\calD = \{ \calP_{11}{} \calP_{12}{}  \calP_{21}{}  \calP_{22}{} \calP_{23}\neq 0\}$. This means that Brans--Dicke system (\ref {eq:DynamicalSystem}) is completely integrable.

\subsection{Jacobi's last multiplier}

Jacobi's last multipliers (JLM)~\cite{23,06,25},  i.e.  functions  $\calM=\calM(P,H,\mitPhi)\neq0$ which obey
		\begin{equation}
\nabla{\cdot}(\calM f)=0
		\label{eq:divMfa}
		\end{equation}
are helpful, in the study of dissipative systems like  (\ref {eq:DynamicalSystem}).
There are  simple relations between Darboux polynomials and Jacobi's last multipliers. For the dynamic system~(\ref {eq:DynamicalSystem}) satisfying condition~(\ref{eq:divMfa}) we get three JLM's
		\begin{equation}
\calM_1=\calP_{11}^{-3}\calP_{12},~~\calM_2=\calP_{11}^{-1}\calP_{12}^{-1}\calP_{21}^{-1},~~\calM_3=\calP_{11}^{-9}\calP_{12}^9
\left[\frac{\calP_{22}}{\calP_{23}}\right]^{-\beta}.
		\label{eq:divMfb}
		\end{equation}
These multipliers build two independent first integrals of Brans--Dicke dynamical system (\ref{eq:DynamicalSystem})
		\begin{equation}
\calI_1=\frac{\calM_1}{\calM_2}~~\mbox{and}~~\calI_2=\frac{\calM_1}{\calM_3}.
		\label{eq:divMfc}
		\end{equation}
This method is an alternative to that shown in subsection \ref{subsec:Darbouxintegrals}.

\subsection{Algebraic invariants}

Irreducible sets of Darboux polynomials are those algebraic invariant solutions which in principle can not be deduced from the first integrals of the dynamical system. Typically, they require a~separate investigation.

 Invariant solutions for cosmological dynamical systems have been investigated in papers~\cite{24,26}. The necessary condition for the existence of invariant sets of the  Brans--Dicke system is the solution to  $\calD' = \{ \calP_{11}{} \calP_{12}{}  \calP_{21}{}  \calP_{22}{} \calP_{23}= 0\}$. The set of solutions to  $\calD'$ can be splitted into invariant subsets

\subitem{$\diamond$}five cases with one zero polynomial Darboux:\\
\vspace*{-1.5ex}
$$
		\begin{array}{lr}
1.1&\{\calP_{11}=0,~\calP_{12}\calP_{21}\calP_{22}\calP_{23}\neq0\}\\
1.2&\{\calP_{12}=0,~\calP_{11}\calP_{21}\calP_{22}\calP_{23}\neq0\}\\
1.3&\{\calP_{21}=0,~\calP_{11}\calP_{12}\calP_{22}\calP_{23}\neq0\}\\
1.4&\{\calP_{22}=0,~\calP_{11}\calP_{12}\calP_{21}\calP_{23}\neq0\}\\
1.5&\{\calP_{23}=0,~\calP_{11}\calP_{12}\calP_{21}\calP_{22}\neq0\}
		\end{array}
$$

\subitem{$\diamond$}two cases with two zero Darboux polynomials:\\
\vspace*{-1.5ex}
$$
		\begin{array}{lr}
2.1&\{\calP_{12}=\calP_{21}=0,~\calP_{11}\calP_{22}\calP_{23}\neq0\}\\
2.2&\{\calP_{22}=\calP_{23}=0,~\calP_{11}\calP_{12}\calP_{21}\neq0\}
		\end{array}
$$

\subitem{$\diamond$}two cases with three zero polynomials Darboux:\\
\vspace*{-1.5ex}
$$
		\begin{array}{lr}
3.1&\{\calP_{11}=\calP_{21}=\calP_{22}=0,~\calP_{12}\calP_{23}\neq0\}\\
3.2&\{\calP_{11}=\calP_{21}=\calP_{23}=0,~\calP_{12}\calP_{22}\neq0\}
		\end{array}
$$

\subitem{$\diamond$}one case with four zero polynomials of Darboux:\\
\vspace*{-1.5ex}
$$
		\begin{array}{lr}
4.1&\{\calP_{11}=\calP_{12}=\calP_{22}=\calP_{23}=0,~\calP_{21}\neq0\}
		\end{array}
$$

\subitem{$\diamond$}one case with five zero polynomials of Darboux:\\
\vspace*{-1.5ex}
$$
		\begin{array}{lr}
5.1&\{\calP_{11}=\calP_{12}=\calP_{21}=\calP_{22}=\calP_{23}=0\}
		\end{array}
$$

Tables~\ref{tab:01} and~\ref{tab:02}  in Appendix present the results of research on the above mentioned invariant subsets of the $\calD'$ set. They include reduced dynamic systems, first integrals and some exact solutions.

\section{Final remarks}
\label{sec:08}

The spacetime with maximally symmetric constant-time hypersurfaces (six parameter group of isometries) is defined up to the time dependent scale factor $R(t)$. In Einstein theory of gravity such a~space is governed by the second order differential equation. The evolution take place in two dimensional phase space what realize the classically understood one--degree mechanical freedom.

In Brans--Dicke theory, the spacetime of the same symmetry is governed by the third order differential equation. The evolution runs in three dimensional phase-space, consequently, first integrals have no natural interpretation of the momentum  conservation or the conformal energy conservation. One may encounter here complex interpretational issues, similar to these arising in kinetics of chemical reactions~\cite{27}.

\section*{Acknowledgements}

Z.A. Golda and A. Woszczyna thank the John Templeton Foundation for supporting this work.
Publication supported by the John Templeton Foundation Grant ``Conceptual Problems in Unification Theories'' (No. 60671).

\newpage

\section{Appendix}

\vspace{-4mm}

\setlength{\arraycolsep}{1.4pt}
\begin{table}[h!]
\footnotesize
\caption{Algebraic invariants, reduced systems and first integral.}
\label{tab:01}
\vspace{-5pt}
\begin{center}
\begin{tabular}{ | c || l | l |}
 \hline
No &\multicolumn{1}{c|}{Algebraic invariants \& dynamical system} & \multicolumn{1}{c|}{First integrals}\\
\hline
\hline
1.1&
\multicolumn{1}{c|}{\parbox{1cm}{
\vspace*{-0.8em}
\begin{align*} 
\calP_{11}=0, ~\calP_{12}\calP_{21}\calP_{22}\calP_{23}\neq 0 \Rightarrow P=0
\end{align*}
\vspace*{-1.2em}
$
\begin{cases}
&\hspace{-1em}H'= -H^2+H\mitPhi-\frac13\omega\mitPhi^2\\ 
&\hspace{-1em}\mitPhi'=-(3H+\mitPhi)\mitPhi
\end{cases}
$
}
}
 & 
\parbox{4.5cm}{
\vspace*{-0.5em}
\begin{equation*}
\begin{split}
i_1  &=  \frac{[6H^2+6H\mitPhi-\omega\mitPhi^2]^3}{\mitPhi^2}\times\\
     &\phantom{=} {\times}\left[
 \frac{6H+(3+\sqrt{9+6\omega})\mitPhi}{6H+(3-\sqrt{9+6\omega})\mitPhi}
\right]^\beta\\
\end{split}
\end{equation*}
\vspace*{-0.5em}}\\
\hline
1.2&
\multicolumn{1}{c|}{\parbox{1cm}{
\vspace*{-0.8em}
\begin{align*} 
\calP_{12}=0, ~\calP_{11}\calP_{21}\calP_{22}\calP_{23}\neq 0 \Rightarrow \mitPhi=0
\end{align*}
\vspace*{-1.2em}
$\begin{cases}
&\hspace{-1em}P'= -4P H \\ 
&\hspace{-1em}H'=-\frac{1}{3}\left[P+3H^2\right]
\end{cases}$
\vspace*{0.2em}
}
}
 & 
\parbox{4.5cm}{
\vspace*{-1.0em}
\begin{equation*}
\begin{split}
i_1 & =  \frac{[P-3 H^2]^2}{P}\\
t_0 & =  \frac{3 H}{P-3 H^2}+t
\end{split}
\end{equation*}
\vspace*{-1.0em}
}\\
\hline 
1.3&
\multicolumn{1}{c|}{\parbox{1cm}{
\vspace*{-0.8em}
\begin{align*} 
&\calP_{21}=0, ~\calP_{11}\calP_{12}\calP_{22}\calP_{23}\neq 0 \Rightarrow\\[-3pt]
& \Rightarrow P=
3H^2+3H\mitPhi-\frac12\omega \mitPhi^2
\end{align*}
\vspace*{-1em}
$\begin{cases}
&\hspace{-1em}H'= -2H^2-\frac16\omega\mitPhi^2\\ 
&\hspace{-1em}\mitPhi'=-(3H+\mitPhi)\mitPhi
\end{cases}$
}
}
 & 
\parbox{4.5cm}{
\vspace*{-1.0em}
\begin{equation*}
\begin{split}
i_1 & =  \frac{(\sqrt{9+6\omega}-3)H+\omega\mitPhi}{(\sqrt{9+6\omega}+3)H-\omega\mitPhi}\times\\
&\phantom{=} {\times}\left[
\frac{(6H^2+6H\mitPhi-\omega\mitPhi^2)^3}{\mitPhi^4}
\right]^{1/\beta}\\
\end{split}
\end{equation*}
\vspace*{-1.0em}}
\\
\hline
1.4&
\multicolumn{1}{c|}{\parbox{1cm}{
\vspace*{-0.8em}
\begin{align*} 
&\calP_{22}=0, ~\calP_{11}\calP_{12}\calP_{21}\calP_{23}\neq 0 \Rightarrow\\[-3pt] &\Rightarrow P=-\frac{\sqrt{3+2\omega}}{2\sqrt3}
\left[
6H+(3+\sqrt{9+6\omega})\mitPhi
\right]\mitPhi
\end{align*}
\vspace*{-1em}
$\begin{cases}
&\hspace{-1em}H'= -H^2+\frac16(3+\sqrt{9+6\omega})(2H+\mitPhi)\mitPhi\\ 
&\hspace{-1em}\mitPhi'=-(3H+\mitPhi)\mitPhi
\end{cases}$
}
}
 & 
\parbox{4.5cm}{
\vspace*{-1.0em}
\begin{equation*}
\begin{split}
i_1 & =  \frac{\left[6H+(3+\sqrt{9+6\omega})\mitPhi\right]^3}{\mitPhi}\times\\
&\phantom{=} {\times}\left[
\frac{6H+(3+\sqrt{9+6\omega})\mitPhi}{2H+\mitPhi}
\right]^\beta\\
\end{split}
\end{equation*}
\vspace*{-1.0em}}
\\
\hline
1.5&
\multicolumn{1}{c|}{\parbox{1cm}{
\vspace*{-0.8em}
\begin{align*} 
&\calP_{23}=0,~\calP_{11}\calP_{12}\calP_{21}\calP_{22}\neq 0 \Rightarrow\\[-3pt] 
&\Rightarrow P=\frac{\sqrt{3+2\omega}}{2\sqrt3}
\left[
6H+(3-\sqrt{9+6\omega})\mitPhi
\right]\mitPhi
\end{align*}
\vspace*{-1em}
$\begin{cases}
&\hspace{-1em}H'= -H^2+\frac16(3-\sqrt{9+6\omega})(2H+\mitPhi)\mitPhi\\ 
&\hspace{-1em}\mitPhi'=-(3H+\mitPhi)\mitPhi
\end{cases}$
\vspace*{0.2em}
}
}
 & 
\parbox{4.5cm}{
\vspace*{-1.0em}
\begin{equation*}
\begin{split}
i_1 & =  \frac{\left[6H+(3-\sqrt{9+6\omega})\mitPhi\right]^3}{\mitPhi}\times\\
&\phantom{=} {\times}\left[
\frac{2H+\mitPhi}{6H+(3-\sqrt{9+6\omega})\mitPhi}
\right]^\beta\\
\end{split}
\end{equation*}
\vspace*{-1.0em}}
\\
\hline
\end{tabular}
\end{center}
\end{table}


\vspace{-5mm}
\setlength{\arraycolsep}{1.4pt}
\begin{table}[h!]
\footnotesize
\caption{Algebraic invariants, reduced systems, first integral and exact solutions.}
\label{tab:02}
\vspace{5pt}
\begin{tabular}{|c|| l | l | l |}
 \hline
No &\multicolumn{1}{c|}{Algebraic invariants \& dynamical system} & \multicolumn{1}{c|}{First integrals} &\multicolumn{1}{c|}{Exact solutions}\\
\hline
\hline
2.1&
\multicolumn{1}{c|}{\parbox{1cm}{
\begin{align*} 
	&\calP_{12}=\calP_{21}~=0,~\calP_{11}\calP_{22}\calP_{23}\neq 0 \Rightarrow\\[-3pt]
	& \Rightarrow P=3H^2~\&~\mitPhi=0
\end{align*}
\vspace*{-1.2em}
$\begin{cases}
&\hspace{-1em}H'=-2H^2
\end{cases}$
}
}
 & 
\parbox{3.8cm}{\begin{equation*}
\begin{split}
t_0 & =-\frac{1}{2H}+t
\end{split}
\end{equation*}}
&\parbox{1cm}{
\vspace*{-0.5em}
\begin{align*}
P&=\frac{3}{4(t-t_0)^{2}}\\
H &=\frac{1}{2(t-t_0)}\\
\mitPhi &=0
\end{align*}
\vspace*{-1.5em}}  \\
\hline
2.2&
\multicolumn{1}{c|}{\parbox{1cm}{
\begin{align*} 
	&\calP_{22}=\calP_{23}=0,~\calP_{11}\calP_{12}\calP_{21}\neq 0 \Rightarrow\\[-3pt]
	&\Rightarrow P=-2(3+2\omega)H^2~\&~\mitPhi=-2H
\end{align*}
\vspace*{-1.2em}
$\begin{cases}
&\hspace{-1em}H'=-H^2
\end{cases}$
}
}
 & 
\parbox{3.5cm}{\begin{equation*}
\begin{split}
t_0 & =  -\frac{1}{H}+t
\end{split}
\end{equation*}}
&\parbox{1cm}{
\vspace*{-0.50em}
\begin{align*}
P&=-\frac{2(3+2\omega)}{(t-t_0)^2}\\
H &=\frac{1}{t-t_0}\\
\mitPhi &=-\frac{2}{t-t_0}
\end{align*}
\vspace*{-0.75em}}  \\
\hline
3.1&
\multicolumn{1}{c|}{\parbox{1cm}{
\vspace*{-0.75em}
\begin{align*} 
&\omega\neq-4/3~\&~\omega\neq0;\\[-3pt]
	&\calP_{11}=\calP_{21}=\calP_{22}=0,~\calP_{12}\calP_{23}\neq 0 \Rightarrow\\[-3pt]
	&  \Rightarrow P=0~\&~\mitPhi=\frac{3-\sqrt{9+6\omega}}{\omega}H
\end{align*}
\vspace*{-1em}
$\begin{cases}
&\hspace{-1em}H'=\frac{-3-3\omega+\sqrt{9+6\omega}}{\omega}H^2
\end{cases}$
}
}
 & 
\parbox{3.5cm}{\begin{equation*}
\begin{split}
t_0 & =-\frac{3+3\omega+\sqrt{9+6\omega}}{3(4+3\omega) H} +t
\end{split}
\end{equation*}}
&\parbox{1cm}{
\vspace*{-0.75em}
\begin{align*}
P&=0\\
H &=\frac{3+3\omega+\sqrt{9+6\omega}}{3(4+3\omega)(t-t_0)}\\
\mitPhi &=\frac{1-\sqrt{9+6\omega}}{(4+3\omega)(t-t_0)}
\end{align*}
\vspace*{-0.75em}}  \\
\hline
3.2&
\multicolumn{1}{c|}{\parbox{1cm}{
\vspace*{-0.75em}
\begin{align*} 
&\omega\neq-4/3~\&~\omega\neq0;\\[-3pt]
	&\calP_{11}=\calP_{21}=\calP_{23}=0,~\calP_{12}\calP_{22}\neq 0 \Rightarrow\\[-3pt]
	&P=0~\&~\mitPhi=\frac{3+\sqrt{9+6\omega}}{\omega}H
\end{align*}
\vspace*{-1em}
$\begin{cases}
&\hspace{-1em}H'=-\frac{3+3\omega+\sqrt{9+6\omega}}{\omega}H^2
\end{cases}$
}
}
 & 
\parbox{3.5cm}{\begin{equation*}
\begin{split}
t_0 & =-\frac{3+3\omega-\sqrt{9+6\omega}}{3(4+3\omega)H}+t
\end{split}
\end{equation*}}
&\parbox{1cm}{
\vspace*{-0.75em}
\begin{align*}
P&=0\\
H &=\frac{3+3\omega-\sqrt{9+6\omega}}{3(4+3\omega)(t-t_0)}\\
\mitPhi &=\frac{1+\sqrt{9+6\omega}}{(4+3\omega)(t-t_0)}
\end{align*}
\vspace*{-0.75em}} \\
\hline
4.1&
\multicolumn{1}{c|}{\parbox{1cm}{
\vspace*{-0.75em}
\begin{align*} 
	&\calP_{11}=\calP_{12}=\calP_{22}=\calP_{23}=0,\\
	&\calP_{21}\neq 0 \Rightarrow P=0~\&~\mitPhi=0
\end{align*}
\vspace*{-1em}
$\begin{cases}
&\hspace{-1em}H'=-H^2
\end{cases}$
}
}
 & 
\parbox{3.5cm}{
\begin{equation*}
\begin{split}
t_0 & = -\frac{1}{H}+t
\end{split}
\end{equation*}}
&\parbox{1cm}{
\vspace*{-0.75em}
\begin{align*}
P&=0\\
H &=\frac{1}{t-t_0}\\
\mitPhi &=0
\end{align*}
\vspace*{-1.5em}}  \\
\hline
5.1&
\multicolumn{3}{c|}{\parbox{1cm}{
\vspace*{-0.75em}
\begin{align*} 
\calP_{11}=\calP_{12}=\calP_{21}=\calP_{22}=\calP_{23}= 0 \Rightarrow P=H=\mitPhi=0~\mbox{(the stationary point)}
\end{align*}
\vspace*{-1.25em}}
}
\\
\hline
\end{tabular}
\end{table}


\end{document}